\documentclass[
amsmath,
amssymb,
aps, 
pra,
twocolumn, 
groupedaddress,
nobalancelastpage,
floatfix]
{revtex4-2}
\usepackage{mathtools}
\usepackage[colorlinks]{hyperref}
\usepackage{lipsum}
\usepackage{bbold}
\usepackage{bm}
\usepackage{dsfont}
\usepackage{enumerate}
\usepackage{graphicx}
\usepackage{float}
\usepackage {xcolor}
\usepackage{cleveref}
\usepackage{amsmath}

\DeclareMathOperator{\e}{e}					% Exponential e
\newcommand{\nullarg}{{}\cdot{}}

\DeclarePairedDelimiterX{\mean}[1]{\langle}{\rangle}{
	\ifblank{#1}{\nullarg}{#1}
}
\DeclarePairedDelimiterX{\abs}[1]{\lvert}{\rvert}{
	\ifblank{#1}{\nullarg}{#1}
}
\DeclarePairedDelimiterX{\norm}[1]{\lVert}{\rVert}{
	\ifblank{#1}{\nullarg}{#1}
}
\DeclarePairedDelimiterX{\bra}[1]{\langle}{\rvert}{#1}

\DeclarePairedDelimiterX{\ket}[1]{\lvert}{\rangle}{#1}
\DeclarePairedDelimiterX{\braket}[2]{\langle}{\rangle}{
	\ifblank{#1}{\nullarg}{#1} \delimsize\vert \ifblank{#2}{\nullarg}{#2}
}
\DeclarePairedDelimiterX{\sandwich}[3]{\langle}{\rangle}{
	\ifblank{#1}{\nullarg}{#1} \delimsize\vert \ifblank{#2}{\nullarg}{#2} \delimsize\vert \ifblank{#3}{\nullarg}{#3}
}
\DeclarePairedDelimiterX{\inner}[2]{\langle}{\rangle}{
	\ifblank{#1}{\nullarg}{#1} , \ifblank{#2}{\nullarg}{#2}
}

\begin{document}
	
	\title{Noninteger high-harmonic generation from extended correlated systems}
	
\author{Christian Saugbjerg Lange} \thanks{Christian Saugbjerg Lange and Thomas Hansen contributed equally to this work.}
\author{Thomas Hansen} \thanks{Christian Saugbjerg Lange and Thomas Hansen contributed equally to this work.}
\author{Lars Bojer Madsen} 
\affiliation{Department of Physics and Astronomy, Aarhus University, Ny Munkegade 120, DK-8000 Aarhus C, Denmark}
\date{\today}
\begin{abstract}
The spectra produced by high-harmonic generation (HHG) typically exhibit well-defined peaks at odd integers times the laser frequency. However, in recent investigations of HHG from correlated materials, spectra exhibit signals at noninteger harmonics which do not conform to the well-known symmetry-based selection rules for HHG-spectra. Here, we use the Fermi-Hubbard model to study HHG from a linear chain of atoms. This model allows us to study both the correlated and uncorrelated phases through a specification of the amount of onsite electron-electron repulsion. The presence of signal at noninteger harmonics can be interpreted as originating from the population of multiple Floquet states. We show how this coupling to different Floquet states depends on the characteristics of the driving pulse and the strength of the electron-electron interaction in the system.
\end{abstract}

\maketitle

%%%%% INTRODUCTION %%%%%
\section{Introduction}
In the process of high-harmonic generation (HHG), a medium (atomic gas, molecules, solid) is driven by an intense laser pulse causing highly nonlinear electron dynamics. The result is the emission of electromagnetic radiation in pulses that can be sub-femtoseconds in duration. The attosecond pulses and the HHG process itself enable the study of electrons at their natural timescale \cite{Krausz2009, Kraus2015}. If the HH generating system is inversion symmetric and if the driving pulse is sufficiently long, selection rules are imposed on the HHG spectrum allowing only odd harmonics \cite{Neufeld2019}. This is verified in both atomic gasses \cite{LHuillier1988}, molecules \cite{Lyngaa1996}, and solids \cite{Ghimire2011, Ghimire2019, Goulielmakis2022}. 

In recent years, there has been an increasing interest in HHG from correlated materials, i.e., materials where a beyond mean-field electron-electron repulsion is of significance. These systems cannot be described by a multiband picture but are instead  described by effective models, such as the Fermi-Hubbard model \cite{Essler_Hubbard_book}, which allows for a description of electrons moving in a chain of atoms in terms of effective hopping and onsite electron-electron repulsion parameters. With this model it was found that the electron-electron correlation can lead to an enhanced signal for certain harmonics \cite{Silva2018, Murakami2018, Murakami2018_2, Hansen22, Hansen22_2, Murakami2021, AlShafey2023, Lange2023electroncorrelation, Hansen2023effects_of_lattice_imperfections}. Curiously, however, HHG spectra from these types of systems show noninteger harmonics \cite{Silva2018, Murakami2021, Murakami2018, Murakami2018_2,Hansen22, Hansen22_2, AlShafey2023, Lange2023electroncorrelation, Hansen2023effects_of_lattice_imperfections} which do not obey the symmetry-imposed selection rules \cite{Neufeld2019}. Furthermore, even if these symmetries are broken, signal is still only expected to be found at (both even and odd) integer harmonics. The origin of these noninteger harmonics have been hypothesized to be the presence of correlations \cite{Silva2018} or due to a short pulse or missing dephasing channels \cite{Murakami2021}, or has simply not been discussed \cite{Murakami2018, Murakami2018_2,Hansen22, Hansen22_2, AlShafey2023, Lange2023electroncorrelation, Hansen2023effects_of_lattice_imperfections}. Thus, a satisfactory explanation of the reported signal at noninteger harmonics in HHG-spectra from correlated systems is lacking.

In this work, therefore, we discuss the physical origin of the signal at noninteger harmonics. To do so, we use the prototypical Fermi-Hubbard model \cite{Essler_Hubbard_book}. This model captures aspects of physics relevant to real materials, including cuprates and some high-$T_c$ superconductors \cite{Lee06, Imada98}. Conveniently, this model allows us to treat the electron-electron correlation strength, the Hubbard-$U$, as a tunable parameter which enables us to study the model with various degrees of electron-electron correlations, ranging from the uncorrelated tight-binding limit to the highly correlated Mott-insulating limit, where the electrons are distributed evenly among the sites.

The paper is organized as follows. First, in Sec. \ref{sec:model_and_observables}, the Fermi-Hubbard model is discussed. The results are presented in Sec. \ref{sec:results}. Here, the uncorrelated case, a case with intermediate correlation strength, and finally the highly correlated Mott-insulating phase are investigated. We conclude and summarize our findings in Sec. \ref{sec:summary_and_conclusion}. Throughout this paper atomic units are used unless explicitly stated otherwise.

%%%%% MODEL AND OBSERVABLES %%%%%
\section{Model and observables} \label{sec:model_and_observables}
We study HHG spectra generated by strong-field driving of the Fermi-Hubbard model \cite{Essler_Hubbard_book}. We consider a one-dimensional chain of atoms corresponding to a one-band system at half filling with the Hamiltonian given as
\begin{equation}
	\hat{H}(t) = \hat{H}_{hop}(t) + \hat{H}_U. \label{eq:Fermi_Hubbard_Hamilton_1}
\end{equation}
Here
\begin{align}
	\hat{H}_{hop}(t) &= - t_0 \sum_{j, \mu} \big(\e^{i a A(t)} \hat{c}_{j, \mu}^\dagger \hat{c}_{j+1, \mu} + \text{H.c.} \big),\label{eq:H_hop} \\ 
	\hat{H}_U &= U \sum_j (\hat{c}_{j, \uparrow}^\dagger \hat{c}_{j, \uparrow}) (\hat{c}_{j, \downarrow}^\dagger \hat{c}_{j, \downarrow}),
\end{align}
describes the electron hopping and onsite electron-electron repulsion, respectively. The parameter $t_0$ describes the strength of an electron hop from site $j$ to its neighboring sites $j \pm 1$ (with periodic boundary conditions), $a$ is the lattice spacing, $A(t)$ is the vector potential of the driving laser polarized along the lattice dimension, and $U$ is the onsite electron-electron repulsion strength. The electronic annihilation operator for an electron with spin $\mu \in \{\uparrow, \downarrow\}$ on site $j$ is denoted $\hat{c}_{j, \mu}$ with the creation operator $\hat{c}_{j,\mu}^\dagger$. We treat $U$ as a tunable parameter allowing us to study the system with various degrees of electron-electron correlation. As apparent from Eq. (\ref{eq:H_hop}), the laser-matter interaction is described through the Peierls phase $\exp(i a A(t))$ \cite{Feynman1966, Essler_Hubbard_book}. In the limiting case of vanishing $U$, the system is simply a tight-binding model as the Hamiltonian of the system in Eq. (\ref{eq:Fermi_Hubbard_Hamilton_1}) reduces to Eq. (\ref{eq:H_hop}). For a finite $U\neq0$, however, onsite electron-electron interactions are apparent, making the system more involved.

The current operator for this model is given as \cite{Murakami2021, Hansen22, Lange2023electroncorrelation}

\begin{equation}
	\hat{j}(t) =  ia t_0 \sum_{j, \mu} \big( \e^{i a A(t)} \hat{c}_{j, \mu}^\dagger \hat{c}_{j+1, \mu} - \text{H.c.} \big). \label{eq:current_operator}
\end{equation}
The expectation value of the current is calculated as
\begin{equation}
	j(t) = \bra{\Psi(t)} \hat{j}(t) \ket{\Psi(t)}, \label{eq:current_expectation_value}
\end{equation}
where $\ket{\Psi(t)}$ is the time-dependent wave function evolved according to the time-dependent Schrödinger equation (TDSE)
\begin{equation}
	i \dfrac{\partial}{\partial t} \ket{\Psi(t)} = \hat{H}(t) \ket{\Psi(t)},	 \label{eq:TDSE}
\end{equation}
 with the Hamiltonian given in Eq. (\ref{eq:Fermi_Hubbard_Hamilton_1}). The many-electron state, $\ket{\Psi(t)}$, is expanded in configurations, $\ket{\Phi_I}$, specifying the site occupations through the multi-index $I$, i.e.,
 \begin{equation}
 	\ket{\Psi(t)} = \sum_I C_I(t) \ket{\Phi_I}, \label{eq:Psi_expansion}
 \end{equation}
 where $C_I(t)$ is the time-dependent amplitude to be solved for. 
 
To drive the system, we use a linearly polarized pulse of the form
\begin{equation}
	A(t) = A_0  f(t) \sin \left(\omega_L t+ \dfrac{\pi}{2} \right), \label{eq:vector_potential}
\end{equation}
where $A_0$ is the amplitude of the vector potential and where the dipole approximation is assumed to be valid. The envelope function of the vector potential is given by
\begin{equation}
	f(t) =
	\begin{cases}
		 \sin^2(\dfrac{\omega_L t}{4 N_{on}}),&0<\frac{t}{T} < N_{on} \\
		 1,&N_{on}  <\frac{t}{T} < (N_{on} + N_{pl}) \\
		 \sin^2\bigg(\dfrac{\omega_L (t - N_{pl}T) }{4 N_{on}} \bigg),& (N_{on} + N_{pl}) <\frac{t}{T}
		  \\&<  (2N_{on} + N_{pl}) \\
		  0,& \text{otherwise},  \label{eq:vector_potential_envelope_function}
	\end{cases}
\end{equation}
i.e., as a flat top pulse with a $\sin^2$ ramp. Here $N_{on}$ and $N_{pl}$ are the number of cycles in the ramp and plateau of the pulse, respectively, and $T=  2\pi/\omega_L$ is the period of the laser. We keep the number of cycles on the top constant, specifically $N_{pl} = 10$, and by changing $N_{on}$ we can study the effect of a longer pulse with a less steep ramp.

The observable of interest is the spectrum generated from the HHG process. It is given by
\begin{equation}
	S(\omega) = \lvert\omega^2 ~ \tilde{j}(\omega) \rvert^2, \label{eq:current_fourier_transfor_implicit}
\end{equation}
where $\tilde{j}(\omega)$ is the Fourier-transform of the current in Eq. (\ref{eq:current_expectation_value}).

We note that for a laser pulse of constant amplitude ($f(t) = 1$), the Hamiltonian in Eq. (\ref{eq:Fermi_Hubbard_Hamilton_1}) is periodic with the laser period, i.e., $\hat{H}(t) = \hat{H}(t+T)$. In this limit, the solution to the TDSE [Eq. (\ref{eq:TDSE})] is given by Floquet states on the form \cite{Holthaus_2016, Joachain2011_atoms_in_intense_laser_fields}
\begin{equation}
	\ket{\Psi_j(t)} = \e^{-i \mathcal{E}_j t} \ket{u_j(t)},
\end{equation}
where $\ket{u_{j}(t)} = \ket{u_j(t+T)}$ are $T$-periodic functions and $\mathcal{E}_j$ are the quasienergy levels which are only uniquely defined up to integer multiples of the laser frequency, $\omega_L$. The set of quasienergies, $\{\mathcal{E}_j\}$, and their corresponding Floquet functions are determined by the parameters of the electronic system but also by the amplitude, $A_0$, and frequency, $\omega_L$, of the vector potential \cite{Joachain2011_atoms_in_intense_laser_fields}. 

Furthermore, we note that for a constant vector potential amplitude and for all correlation strengths, the system is symmetric under reflection in space and time translation by half a period. That is $\hat{C}_2 \hat{H}(t) \hat{C}^{-1}_2 = \hat{H}(t)$. Here $\hat{C}_2 = \hat{R}_2 \cdot \hat{\tau}_2$, where $\hat{R}_2$ is reflection of the electronic system and $\hat{\tau}_2$ is a time translation by $T/2$. With this symmetry and if the system only populates a single Floquet state, only odd harmonics are allowed in the spectrum as shown in Ref. \cite{Neufeld2019}.

Here, we investigate a system of $L=10$ sites with periodic boundary conditions with a lattice spacing of $a = 7.5588$ a.u., and $t_0 = 0.0191$ a.u. similar to Refs. \cite{Silva2018, Hansen22, Hansen22_2, Hansen2023effects_of_lattice_imperfections, Lange2023electroncorrelation}. These values are picked to fit those of the cuprate $\text{Sr}_2\text{CuO}_3$ \cite{Tomita2001}. The field has an amplitude of $A_0 = F_0/\omega_L = 0.194$ a.u. with angular frequency $0.005 \text{ a.u.} = 33$ THz. This choice of field strength corresponds to a peak intensity of $F_{0} = 3.3 \times 10^{10}~\text{W}/\text{cm}^2$.
To solve the dynamics of the system driven by the laser governed by the TDSE in Eq. (\ref{eq:TDSE}), we use the Arnoldi-Lancoz algorithm \cite{Park1986, Smyth1998, Guan2007, Frapiccini2014} with a Krylov subspace of dimension $4$. To simplify the numerical calculations, we exploit that the Hamiltonian in Eq. (\ref{eq:Fermi_Hubbard_Hamilton_1}) is invariant under spin-flip of all electrons and under translations of the entire system corresponding to spin-flip symmetry and conservation of total crystal momentum, respectively \cite{Essler_Hubbard_book}. In this manner, the number of configurations entering the expansion in Eq. (\ref{eq:Psi_expansion}) is reduced. In our simulations, we start from a nondegenerate spin symmetric ground state with vanishing total crystal momentum, and as such only states within that subspace are needed in the basis. All results have been checked for convergence.

%%%%% RESULTS %%%%%
\section{Results and discussion} \label{sec:results}
\begin{figure}[t]
	\centering
	\includegraphics[width=1\linewidth]{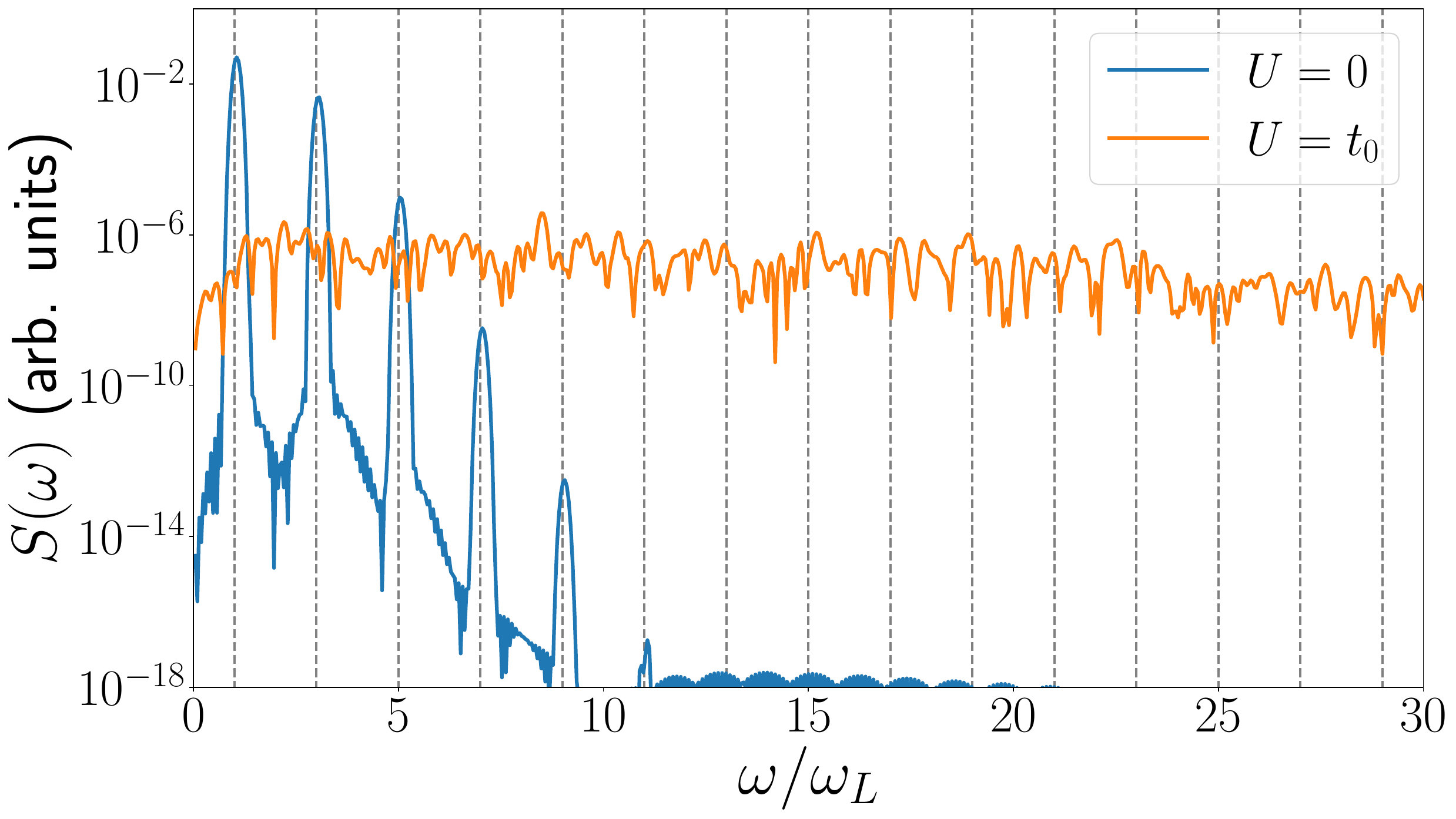}
	\caption{Spectrum for an uncorrelated system $(U=0$) and a correlated system ($U=t_0$). We note how the presence of correlations completely changes the spectrum. Both spectra have been obtained with an identical pulse with $N_{on}=3$ [see Eq. (\ref{eq:vector_potential_envelope_function})] and parameters specified in the text. The gray vertical dashed lines indicate odd harmonics to guide the eye.}
	\label{fig1}
\end{figure}

In this section, we show the effects of correlations and pulse characteristics on the generated HHG spectra. The results presented will be followed by a discussion based on Floquet theory to explain the observations. In Fig. \ref{fig1} the spectra for an uncorrelated system ($U=0$) and for a system with an intermediate correlation strength of $U=t_0$ are shown. We note how the clear peaks in the uncorrelated phase at lower odd harmonics ($\omega /  \omega_L  \leq 11$) completely disappear when correlations are introduced. Furthermore, we note that correlations enhance the spectrum and extend it to much higher harmonics without any clear peaks, indicating that the presence of correlations drastically changes the underlying physics that generates the spectrum. While this enhancement and extension of the spectrum has already been observed and is well understood in the Mott-insulating phase \cite{Silva2018, Murakami2021,Hansen22, Hansen22_2, Murakami2018}, a satisfactory explanation of the absence of well-defined peaks is still missing. We will in this section first discuss the uncorrelated case before investigating the case with an intermediate correlation strength of $U=t_0$. At the end of the section also the Mott-insulating phase ($U=10 t_0$) will be investigated.

\subsection{Uncorrelated phase}
When $U=0$ the Hamiltonian reduces to a simple tight-binding model with $\hat{H}(t)=\hat{H}_{hop}(t)$ in Eq. (\ref{eq:H_hop}). This case corresponds to a one-band model, where only intraband harmonics can be generated. The importance of this limit in understanding experimental data is well-documented \cite{Ghimire2011, Ghimire2012, Luu2015}. It can be shown that $\hat{H}_{hop}(t)$ commutes with itself at all times \cite{Lange2023electroncorrelation}. It then follows that the time evolution operator in this limit is given as $\hat{\mathcal{U}}(t,t_i) = \exp(-i \int_{t_i}^{t} \hat{H}_{hop}(t')dt')$. Taking the many-electron initial state, $\ket{\Psi(t_i)}$, to be an eigenstate of the field-free system, the state in the uncorrelated phase at any time is therefore given as
\begin{equation}
	\ket{\Psi(t)} =\hat{\mathcal{U}}(t,t_i) \ket{\Psi(t_i)} =  \e^{-i \int_{t_i}^{t}  E(t') dt'} \ket{\Psi(t_i)}, \label{eq:psi_t_uncorrelated}
\end{equation}
where $E(t)$ is the time-dependent energy and $t_i$ is the initial time at which interactions between the laser and electronic system are turned on. It is seen from Eq. (\ref{eq:psi_t_uncorrelated}) that the time evolution is simply accounted for by a time-dependent phase applied to the initial state. As such, only a single many-electron Floquet state is populated at all times.

To re-express Eq. (\ref{eq:psi_t_uncorrelated}) using Floquet theory, we define
\begin{equation}
	\mathcal{E} = \dfrac{1}{T} \int_{t_i}^{t_i + T} E(t') dt', \label{eq:quasi_energy_uncorrelated}
\end{equation}
which enables us to rewrite the phase in Eq. (\ref{eq:psi_t_uncorrelated}) as 
\begin{equation}
	\e^{-i \int_{t_i}^t [E(t') - \mathcal{E} + \mathcal{E}] dt'} = \e^{-i \mathcal{E}(t-t_i)} \e^{-i \int_{t_i}^t [E(t')- \mathcal{E}]dt'},
\end{equation}
which by inserting into Eq. (\ref{eq:psi_t_uncorrelated}) yields
\begin{equation}
	\ket{\Psi(t)} = \e^{-i \mathcal{E}(t-t_i)}  \ket{u(t)}, \label{eq:uncorrelated_Floquet_state}
\end{equation}
where we have defined $\ket{u(t)} =  \e^{-i \int_{t_i}^t [E(t')- \mathcal{E}]dt'} \ket{\Psi(t_i)}$. We note that in the limit of periodic driving, i.e., when $E(t) = E(t+T)$ the state $\ket{u(t)} = \ket{u(t+T)}$ is periodic. Equation (\ref{eq:uncorrelated_Floquet_state}) expresses the formal solution to the many-electron Hamiltonian [Eq. (\ref{eq:Fermi_Hubbard_Hamilton_1})] in the $U=0$ case in a Floquet picture, and since only a single Floquet state is populated [Eq. (\ref{eq:uncorrelated_Floquet_state})], only odd harmonics are found in the spectrum \cite{Neufeld2019}. This conclusion is also reached by noting that the expectation value of the current [Eq. (\ref{eq:current_expectation_value})] is in this case simply
\begin{equation}
	\langle \hat{j}(t) \rangle  = \bra{u(t)} \hat{j}(t) \ket{u(t)}, \label{eq:current_expectation_value_floquet_uncorrelated}
\end{equation}
which only contains frequency components determined by $A(t)$. The sharpness of the peaks is only limited by the integration limits of the Fourier transform involved in obtaining the spectra [Eq. (\ref{eq:current_fourier_transfor_implicit})] and is hence determined by the bandwidth of the driving pulse. We note, that it is not necessary to investigate the uncorrelated phase from the perspective of Floquet theory to conclude that only odd harmonics are visible in the spectrum. This can also be shown explicitly algebraically [see App. \ref{App:Exact_calculation_of_uncorrelated_phase}].

\begin{figure}[t!]
	\centering
	\includegraphics[width=1\linewidth]{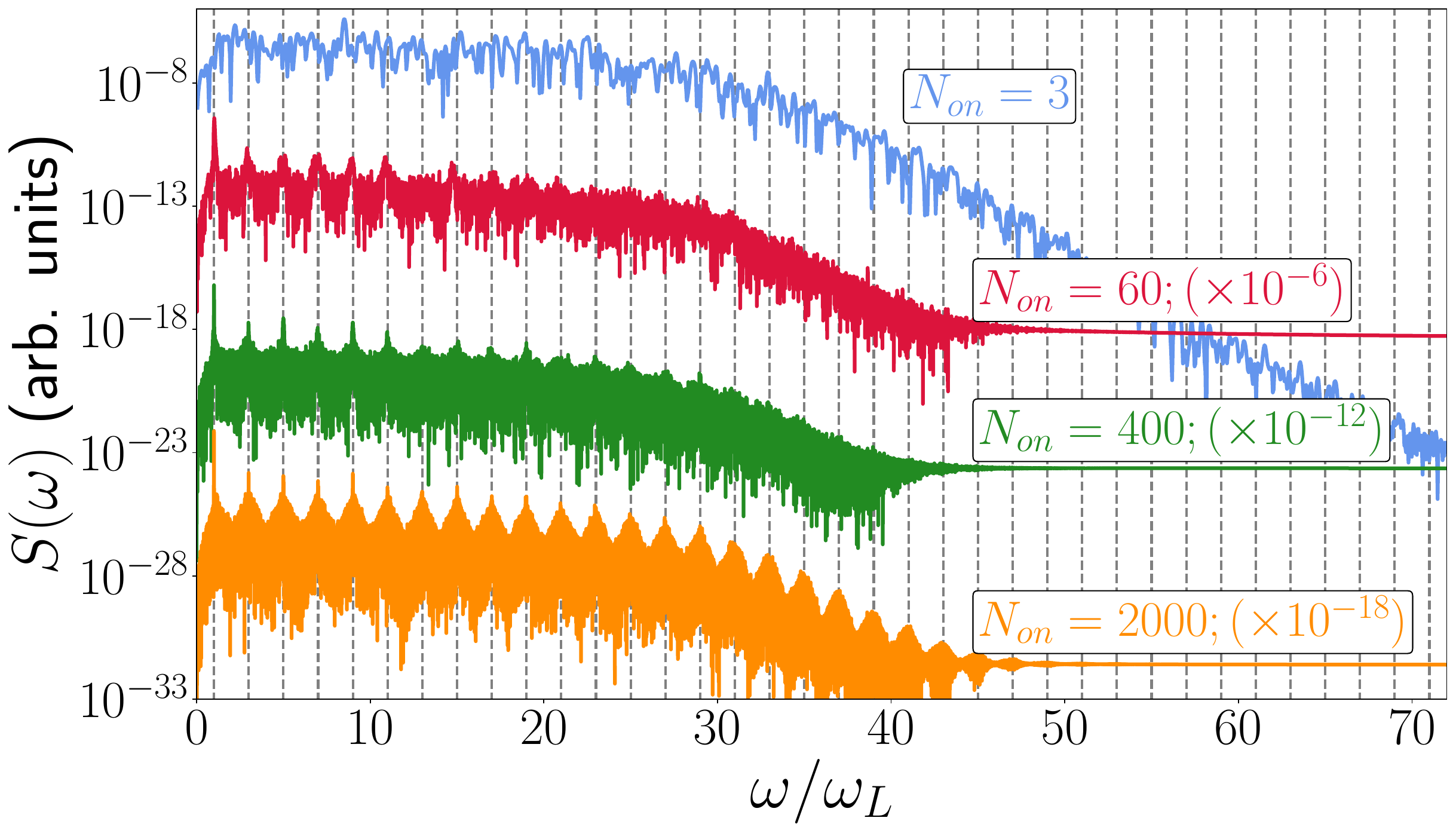}
	\caption{Spectra for a correlated system with $U=t_0$ for turn-on cycles $N_{on} = 3, 60, 400, 2000$ [see Eq. (\ref{eq:vector_potential_envelope_function})]. We note how peaks at odd harmonics become clearer with a more adiabatic (longer) turn-on of the pulse. The numbers in the parenthesis in the inserts indicate that the spectra are scaled for visual clarity.}
	\label{fig2}
\end{figure}

\subsection{Noninteger HHG in a correlated system}
When we introduce correlations to the system, the spectrum changes completely, as seen in Fig. \ref{fig1}. For $U\neq0$ no simple analytical expression for the wave function exists and we rely on numerical simulations. We first use a moderate electron-electron correlation strength of $U=t_0$. To study the effect of the pulse ramp and pulse length, we will in the remainder of this work consider pulses with various ramp lengths, starting from a typical ramp extending to an extremely long pulse with a highly adiabatic ramp by using $N_{on} = 3, 60, 400, 2000$ in Eq. (\ref{eq:vector_potential_envelope_function}).

Naively, one would think that the spectrum presented in Fig. \ref{fig1} does not show clear harmonics due to a short pulse ($N_{on} = 3$). However, this is only partially true as seen in Fig. \ref{fig2} where spectra for the various ramp lengths are shown. Here, the spectra have been scaled for visual clarity as indicated by the number in the parenthesis in the inserts to the right. To obtain a well-resolved spectrum, a window function of $\cos^8$ has been used in all figures for the case of a very short ramp of $N_{on}=3$. This is not necessary for  $N_{on} = 60, 400, 2000$.  As seen in Fig. \ref{fig2}, the signal at odd harmonics becomes clearer with increasing pulse length. Surprisingly, however, we see how the peaks are accompanied by a broadening which does not disappear for a longer ramp though the spectrum becomes more regular. This behavior is different from the uncorrelated case where only sharp peaks are found at the odd harmonics without such a broadening [Fig. \ref{fig1}]. As such, the presence of correlations introduces new phenomena not found in uncorrelated systems.

We will now explain the origin of the signal at noninteger harmonics. As discussed below, the presence of these indicates that, despite the extremely adiabatic turn-on of the laser, the system is still populated by more than a single Floquet state. Different from the uncorrelated case, the presence of correlations now allows coupling between (Floquet) states during the dynamics. In the present case where the amplitude of the vector potential changes with time, the set of quasienergies and corresponding Floquet functions depend on the instantaneous field amplitude, i.e., $\mathcal{E}(t) = \mathcal{E}(A_0(t))$ \cite{Joachain2011_atoms_in_intense_laser_fields}, where we for simplicity define $A_0(t) = A_0 f(t)$ with $f(t)$ given in Eq. (\ref{eq:vector_potential_envelope_function}).  We may formally expand a general state, $\ket{\Psi(t)}$, in terms of Floquet states
\begin{align}
	\ket{\Psi (t)} =  \sum_{j} c_j(t) \e^{-i \int_{t_i}^t \mathcal{E}_j(t') dt'} \ket{u_{\mathcal{E}_j(t)} (t)}.   \label{eq:Floquet_state_most_general}
\end{align}
Here $c_j(t)$ is a time-dependent expansion coefficient to account for couplings between Floquet states. In the absence of couplings $c_j(t) = \delta_{j,i}$, where $i$ denotes the Floquet state initially populated at $t=t_i$. We note that Eq. (\ref{eq:Floquet_state_most_general}) reduces to the form of Eq. (\ref{eq:uncorrelated_Floquet_state}) for a constant field amplitude and no couplings. The time integrals of the quasienergies in Eq. (\ref{eq:Floquet_state_most_general}) reflect that these quasienergies now depend on time through the time-dependence of the field.

The expectation value of the current [Eq. (\ref{eq:current_expectation_value})] for the state in Eq. (\ref{eq:Floquet_state_most_general}) is readily calculated as

\begin{align}
	\langle \hat{j}(t) \rangle &= \sum_{i,j} c_i^*(t) c_j(t) \e^{-i \int_{t_i}^t [\mathcal{E}_j(t') - \mathcal{E}_i(t')]dt'} \nonumber \\
	&\qquad \quad \times \bra{u_{\mathcal{E}_i(t)} (t)} \hat{j}(t) \ket{u_{\mathcal{E}_j(t)} (t)}. \label{eq:current_expectation_value_general_floquet}
\end{align}
Equation (\ref{eq:current_expectation_value_general_floquet}) looks vastly different than if only a single Floquet state is populated as in Eq. (\ref{eq:current_expectation_value_floquet_uncorrelated}). In particular, we see that the phase $\exp{(-i \int_{t_i}^t [\mathcal{E}_j(t') - \mathcal{E}_i(t')]dt')}$ contains frequency components which are in general not a multiple of the laser frequency. That is, if more than a single Floquet state is populated, this phase will generate noninteger harmonics in the spectrum. These are the so-called hyper-Raman lines previously found in HHG studies in atomic systems \cite{Millack1993, Corso1997, DiPiazza2001, Moiseyev2003, Bloch_2019}.

\begin{figure}
	\centering
	\includegraphics[width=1\linewidth]{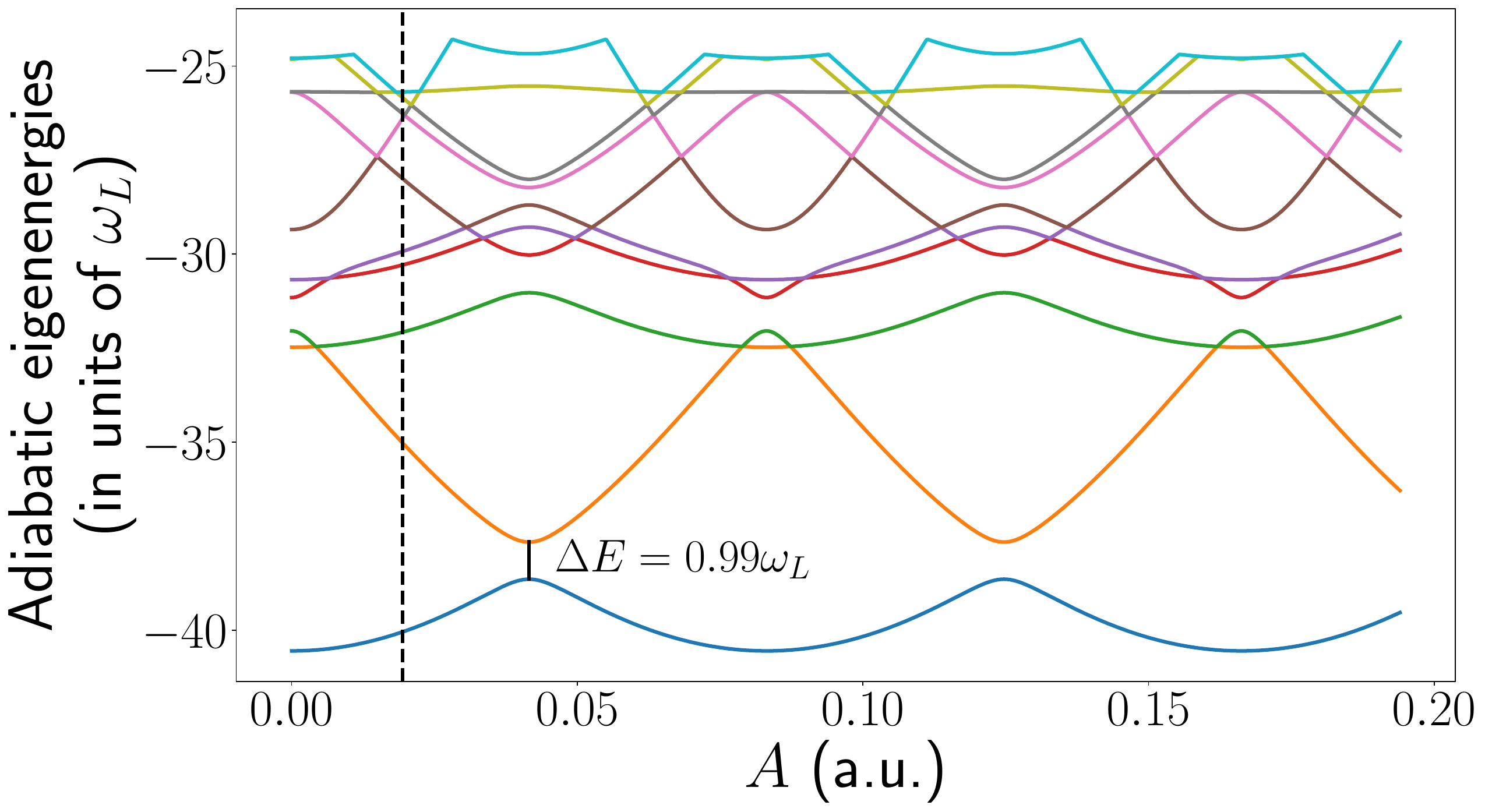}
	\caption{Adiabatic eigenenergies as a function of the magnitude of the vector potential with an electron-electron correlation strength of $U=t_0$ and parameters specified in the main text. We note that for $A \approx 0.04$ a.u. the ground state is at a nonadiabatic one-photon resonance with the first excited state. The dashed vertical line indicates the reduced $A_0'$ value considered (see text).}
	\label{fig3}
\end{figure}

Based on the above discussion and the results presented in Fig. \ref{fig2}, it seems likely that the system indeed populates more than a single Floquet state throughout the dynamics. The population of multiple Floquet states can be explained by possible resonances in the quasienergy spectrum. However, the size of the system and the low frequency of the laser impede a numerical diagonalization of the full Floquet Hamiltonian to obtain the entire spectrum of quasienergies. Instead, we gain insight from the adiabatic eigenenergies which are obtained by diagonalizing $\hat{H}(t)$ in Eq. (\ref{eq:Fermi_Hubbard_Hamilton_1}) for a fixed time and which can be related to the Floquet quasienergies through a perturbative expansion of the latter in even powers of the small laser frequency $\omega_L^{2n}$, $(n=0,1,2,..)$ \cite{Joachain2011_atoms_in_intense_laser_fields}. Figure \ref{fig3} shows the adiabatic eigenenergies for vector potential amplitudes up to the $A_0$ value of $A_0 = 0.194$ a.u. used in the TDSE simulations generating the spectra in Figs. \ref{fig1}, \ref{fig2}, and \ref{fig6}. We note how the gap in energies changes and especially that a one-photon resonance occurs between the two lowest-lying states at around $A \approx 0.04$ a.u., indicating a strong nonadiabatic coupling.

\begin{figure}
 	\centering
 	\includegraphics[width=1\linewidth]{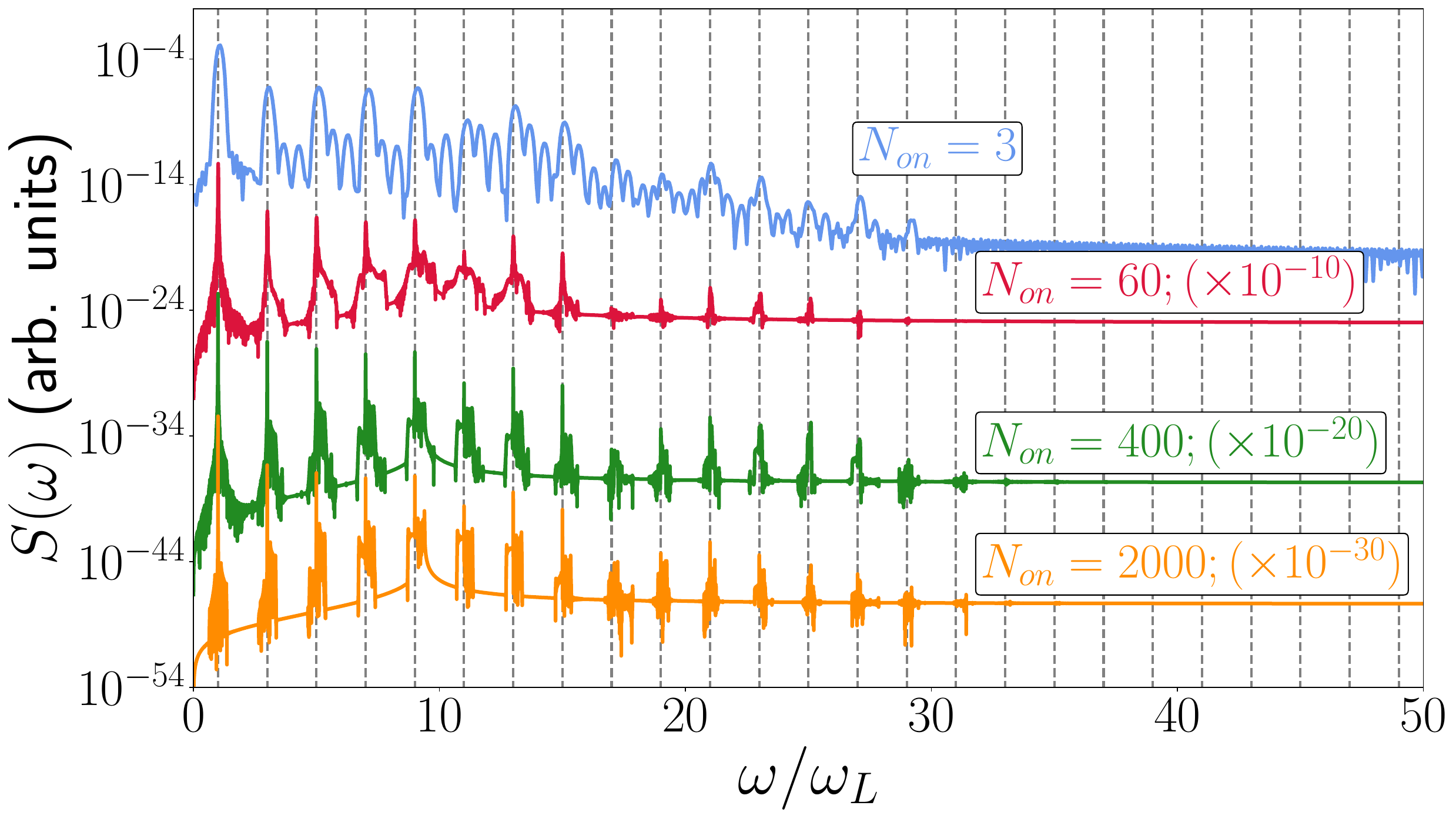}
 	\caption{Spectra for a correlated system with $U=t_0$. The same parameters as in Fig. \ref{fig2} have been used except for a smaller vector potential amplitude of $A_0' = 0.1 A_0 = 0.0194$ a.u.. We note that clear peaks are found at odd harmonics with a much smaller broadening when compared to Fig. \ref{fig2}.}
 	\label{fig4}
 \end{figure}

Based on this observation, we now consider the results of a TDSE simulation for the same system driven with a weaker vector potential amplitude of $A_0' = 0.1 A_0$. This is done in order to be off-resonant with the one-photon transition between the two lowest-lying states [see Fig. \ref{fig3}]. The spectra for the system driven with a vector potential of amplitude $A_0'$ are seen in Fig. \ref{fig4}. Here we clearly see that for all the values of $N_{on}$ considered, peaks are found at odd harmonics with increasing sharpness for longer pulse lengths. We ascribe this finding to the fact that the field strength remains far to the left of the resonance seen in Fig. \ref{fig3}, prohibiting many Floquet states from becoming populated. Furthermore, we also note that though the peaks in Fig. \ref{fig4} are much narrower when compared to the spectra in Fig. \ref{fig2} they still have finite widths even for the longer pulses, indicating that several Floquet states are populated. However, we emphasize that the signal seen at odd harmonics of the laser frequency dominates the noninteger signal by orders of magnitude.

The origin of noninteger HHG in correlated systems can hence be ascribed to the population of multiple Floquet states due to resonances in the Floquet quasienergy spectrum. In the uncorrelated case, only the pulse length dictates the resolution of the peaks at odd harmonics, as only a single Floquet state is populated throughout the dynamics. In the correlated case, not only the length and the ramp of the pulse is of importance but also its amplitude plays a role. This is due to the fact, that when correlations are introduced, couplings are allowed and other Floquet states can become populated. The strength of this coupling depends both on the vector potential amplitude of the driving laser [compare Figs. \ref{fig2} and \ref{fig4}] and on the degree of correlations in the target system. These findings can be used to study the quasienergy levels of a correlated system. In particular, the HHG spectrum can be used to study the presence of such resonances throughout the dynamics by varying the maximum amplitude of the vector potential.

\subsection{Noninteger HHG in Mott-insulators}
We now consider a system with an electron-electron correlation strength of $U=10 t_0$ which places the system in the Mott-insulating phase. Several works have observed noninteger harmonics and a general irregular spectrum above the so-called Mott-gap \cite{Silva2018, Murakami2021, Murakami2018, Hansen22, Hansen22_2, AlShafey2023, Lange2023electroncorrelation}. As mentioned in the introduction, the presence of noninteger HHG has been hypothesized to be due to missing dephasing channels and a short pulse length \cite{Murakami2021} or due to the presence of correlations \cite{Silva2018}, and some works have simply not discussed this aspect of the HHG-spectrum \cite{Murakami2018, Hansen22, Hansen22_2, AlShafey2023, Lange2023electroncorrelation}. In view of the symmetry-based constraints for the present system with $\hat{C}_2$-symmetry allowing only odd harmonics in the long pulse limit \cite{Neufeld2019}, it seems relevant to provide an explanation of the origin of the occurrence of noninteger HHG.

\begin{figure}
	\centering
	\includegraphics[width=1\linewidth]{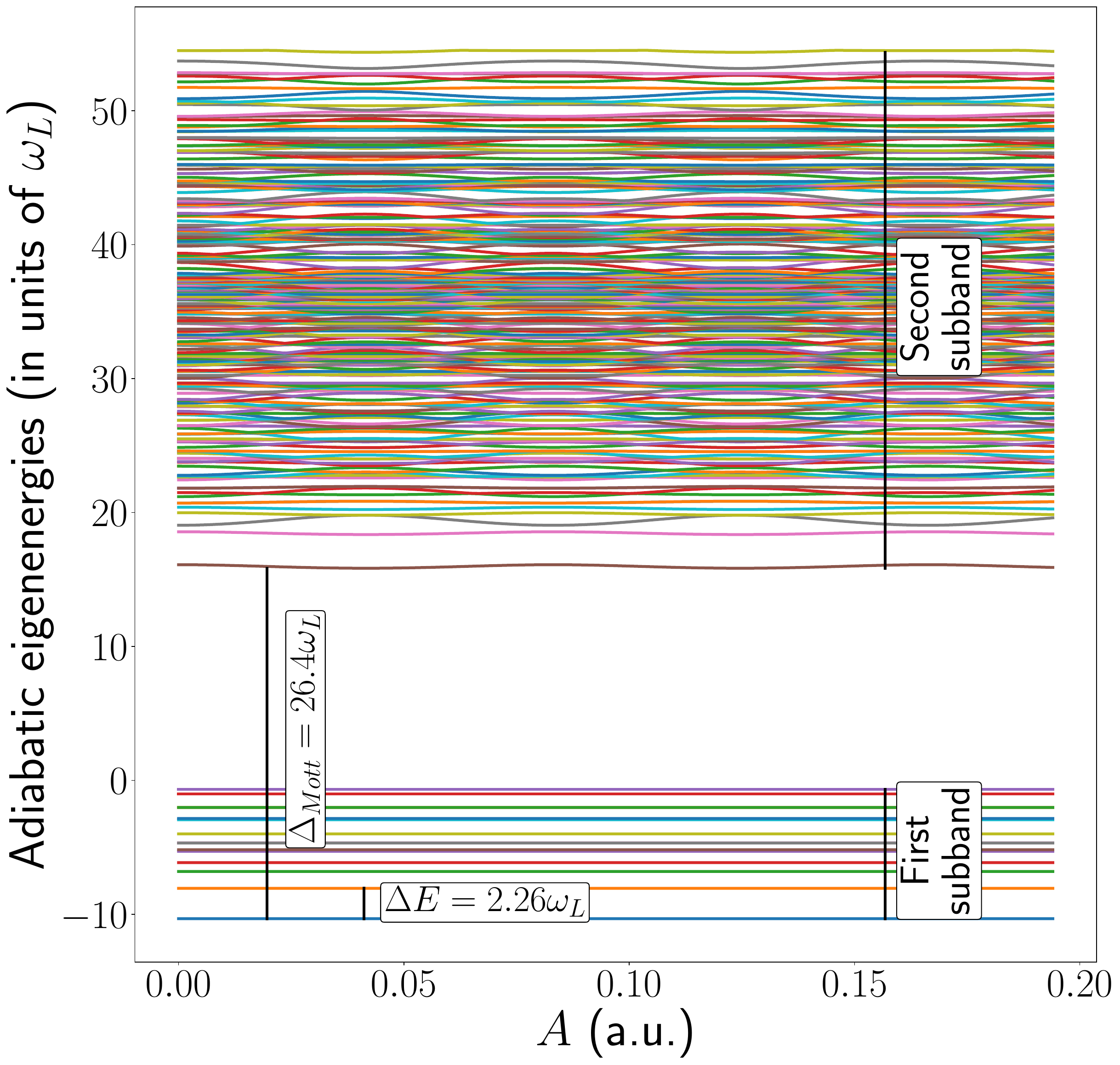}
	\caption{Adiabatic eigenenergies as a function of the magnitude of the vector potential with an electron-electron correlation strength of $U=10 t_0$ and parameters specified in the main text. We note that in this Mott-insulating limit, $\Delta_{Mott}$ is largely independent of the strength of the electric field.}
	\label{fig5}
\end{figure}

Before going into the discussion about HHG in this type of system, we first introduce the relevant physical quantities. In the Mott-insulating limit, the eigenenergies of the field-free system split into several Hubbard subbands [see Fig. \ref{fig5}]. In the first subband, which contains the lowest-energy eigenstates including the ground state, the system is largely dominated by configurations with only a single electron on each site as it is energetically expensive to have two electrons on the same site due to the large value of the Hubbard-$U$. Similarly, the states in the second subband are dominated by configurations that contain a single double-occupancy of electrons on a site. Such a double occupancy is called a doublon and can be considered a quasiparticle. This doublon is accompanied by an empty site called a holon quasiparticle \cite{Murakami2018, Murakami2021, Hansen22_2, Lange2023electroncorrelation, Hansen2023effects_of_lattice_imperfections}. These two lowest subbands are the relevant ones in the present work. Also of relevance is the energy difference between the ground state in the first subband and the lowest state in the second subband described by $\Delta_{Mott}$ called the Mott-gap [see Fig. \ref{fig5}]. This separation gives rise to two different kinds of currents: the intra- and inter-subband currents \cite{Murakami2018, Murakami2021} reminiscent of intra- and inter-band currents for bandgap materials \cite{Ghimire2011, Ghimire2019, Goulielmakis2022}. The intra-subband current originates from the propagation of the states with a single doublon-holon pair within the second subband which results in harmonics with energies below the Mott-gap. The inter-subband current originates from the recombination of a doublon-holon pair. Here transition occurs from the second subband to the first subband resulting in the emission of a harmonic with energy around or above $\Delta_{Mott}$ \cite{Murakami2018, Murakami2021}. In the present work $\Delta_{Mott} = 26.4 \omega_L$. We note that the current operator in Eq. (\ref{eq:current_operator}) involves moving a single electron which will create a single doublon-holon pair when applied to a state within the first subband. As all states within the first subband contain virtually no doublon-holon pairs, the matrix element, $\bra{\Psi_i} \hat{j}(t)\ket{\Psi_j}$ with $\ket{\Psi_{i,j}}$ being states within the first subband, is insignificant and there will be virtually no current contribution from couplings within the first subband, see Fig. \ref{fig5}. Further discussions about the Mott-insulating phase in the presence of intense laser pulses can be found, e.g., in Refs. \cite{Silva2018, Murakami2018, Murakami2018_2, Murakami2021, Hansen22_2, Oka2012}.

\begin{figure}
	\centering
	\includegraphics[width=1\linewidth]{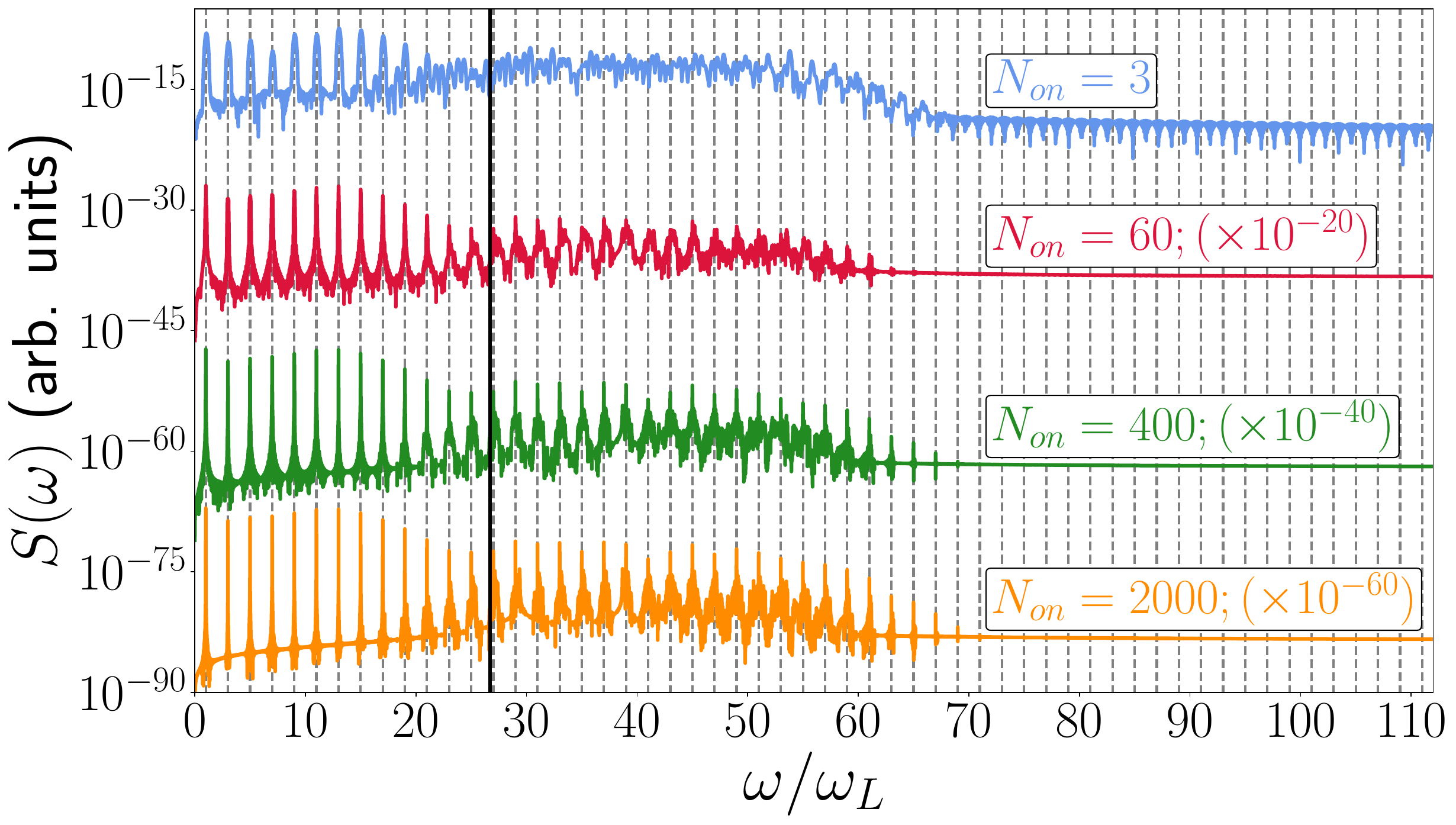}
	\caption{Spectra for the Mott insulating phase with $U=10t_0$ for various pulse lengths and parameters specified in the main text. Above the Mott-gap (solid vertical line) clearer peaks are seen for longer pulse lengths.}
	\label{fig6}
\end{figure}

The spectra obtained for the Mott-insulating phase are shown in Fig. \ref{fig6}. Here, clear peaks at odd harmonics in the intra-subband region with energies below $\Delta_{Mott}$ are seen for all considered pulse lengths. On the contrary, the peaks in the inter-subband region above the Mott-gap become more visible for longer ramp times. We further note, that the broadening of the peaks for $U=t_0$ found in the spectra in Fig. \ref{fig2} is not similarly found in Fig. \ref{fig5}. This indicates that no resonances between states occur with the varying strength of the vector potential in the Mott-insulating phase. Indeed, this is consistent with the adiabatic eigenenergies for the Mott-insulating phase shown in Fig. \ref{fig5}, where we see that the eigenenergies are virtually constant for a varying strength of the vector potential. In other words, the coupling between states does not change with a varying field strength prohibiting the same kind of field control of the dynamics as in the case of $U=t_0$. The possible population of multiple Floquet sates will thus not change notably with a varying amplitude of the vector potential or with a longer pulse, the latter of which is also testified in Fig. \ref{fig6}. It is worth pointing out that it is primarily the peaks in the interband region that are broadened. This is due to the fact that though the dynamics populate multiple Floquet states, the states with a higher energy than the lowest state in the second subband in Fig. \ref{fig5} have relatively little population. Consequently, the contribution to the current from terms which involve only states with a higher energy than the lowest state in the second subband [Fig. \ref{fig5}] is negligible when compared to terms which involve the lowest state in the second subband or states in the first subband as these are in general more populated. The off-diagonal terms contributing to the current in Eq. (\ref{eq:current_expectation_value_general_floquet}) will hence only have a phase with a difference in quasienergy at around or above $\Delta_{Mott}$ which broadens the signal only in the interband region of the spectrum. Similarly to the spectra in Fig. \ref{fig4}, the spectra in Fig. \ref{fig6} become dominated by the signal at odd harmonics by orders of magnitude when compared to the signal at noninteger harmonics with increasing ramp time.

%%%%% SUMMARY AND CONCLUSION %%%%%
\section{Summary and conclusion} \label{sec:summary_and_conclusion}
In this work, we investigated the origin of noninteger harmonics found in spectra modelling the response of correlated solids. We employed the Fermi-Hubbard model and varied the Hubbard $U$, to analyze both the uncorrelated case and the correlated case. In the correlated case we investigated both an intermediate and high degree of electron-electron correlations, the latter known as the Mott-insulating phase. In the uncorrelated case, only odd harmonics were found in the spectrum, consistent with the population of a single Floquet state throughout the dynamics. Conversely, in the correlated phase with moderate electron-electron correlation $(U=t_0)$, even with prolonged ramps, signal appeared at frequencies not an odd multiple of the laser frequency. This latter signal originates from the concurrent population of multiple Floquet states during the dynamics and is not at odds with the strong symmetry-based argument allowing only odd harmonics as this assumes only a single Floquet state to be populated \cite{Neufeld2019}. The varying amplitude of the vector potential causes states to become resonant, allowing multiple Floquet states to couple. Notably, reducing the maximum amplitude of the vector potential mitigated this resonance, resulting in predominantly odd harmonics with minimal broadening, indicating fewer populated Floquet states.
The duration and bandwidth of the pulse crucially determine the resolution of the HHG peaks in the spectrum. However, changes in the vector potential amplitude via the pulse envelope can lead to resonances between states, populating multiple Floquet states which will generate noninteger harmonic signal. Similar observations hold true for the Mott-insulating phase ($U=10t_0$) where quasienergy levels remain largely unaffected by the magnitude of the vector potential amplitude. Thus, a longer pulse suffices to resolve the HHG peaks at odd harmonics in the interband regime of the spectrum. These findings are similar to the so-called hyper-Raman lines predicted \cite{Millack1993, Corso1997, DiPiazza2001, Moiseyev2003} and recently observed \cite{Bloch_2019} in studies of HHG in atoms, now   reemerging in materials when accounting for beyond mean-field electron-electron correlations. However, the large energy gap in atoms between ground and excited states made it difficult  to observe these noninteger spectral features in atomic HHG experiments and necessitated a two-color approach for their observation \cite{Bloch_2019}. In contrast, the energy difference between states in the present model can be much smaller, which could lead one to expect that experimental observations and potential technological applications of noninteger HHG are feasible in correlated materials.

\begin{acknowledgments}
	This work was supported by the Independent Research Fund Denmark (Grant No. 1026-00040B) and by the Danish National Research Foundation through the Center of Excellence for Complex
	Quantum Systems (Grant Agreement No. DNRF156).
\end{acknowledgments}

\appendix

\section{Explicit calculation of uncorrelated phase} \label{App:Exact_calculation_of_uncorrelated_phase}

In this appendix, we show by algebraic derivations, without the use of Floquet theory, that the uncorrelated phase ($U=0$) yields only odd harmonics. The state of the system at all times is given in Eq. (\ref{eq:psi_t_uncorrelated}) in the main text and is restated here for convenience
\begin{equation}
	\ket{\Psi(t)} = \e^{-i \int_{t_i}^{t} E(t') dt'} \ket{\Psi(t_i)}.
\end{equation}
We start by transforming the Fermionic creation operator from a site-specific description (site index $j$) to a crystal momentum-specific description (crystal momenta $q$)
\begin{equation}
	\hat{c}_{j, \mu}^\dagger = \dfrac{1}{\sqrt{L}} \sum_q \e^{-i R_j q} \hat{c}_{q, \mu}^\dagger \label{eq:c_j_transformation}
\end{equation} 
and similarly for the annihilation operator. Here $R_j = j \cdot a$ is the position of the $j$'th site with $a$ being the lattice constant. The transformation in Eq. (\ref{eq:c_j_transformation}) allows us to describe the system in the spatially delocalized Bloch basis instead of the localized Wannier basis underlying the representations in Eqs. (\ref{eq:Fermi_Hubbard_Hamilton_1})-(\ref{eq:current_operator}).  Inserting Eq. (\ref{eq:c_j_transformation}) into the Hamiltonian in Eq. (\ref{eq:H_hop}) yields
\begin{align}
	\hat{H}_{hop}(t) = -2 t_0 \sum_{q, \mu} \cos((q + A(t))a) \hat{c}_{q,\mu}^\dagger \hat{c}_{q,\mu}, \label{eq:hamiltonian_k_space}
\end{align}
and the current operator in Eq.(\ref{eq:current_operator}) likewise transforms as
\begin{align}
	\hat{j}(t) = 2 t_0 a \sum_{q, \mu} \sin((q + A(t))a) \hat{c}_{q,\mu}^\dagger \hat{c}_{q,\mu}. \label{eq:current_k_space}
\end{align}
From Eqs. (\ref{eq:hamiltonian_k_space}) and (\ref{eq:current_k_space}) it is clearly seen that the system is diagonal in $q$-space. Further, we note from Eq. (\ref{eq:hamiltonian_k_space}), that the ground state has a symmetric distribution of electron crystal momenta around $q=0$. 

The expectation value of the current is calculated using Eq. (\ref{eq:current_k_space})
\begin{align}
	j(t)= \langle \hat{j}(t) \rangle &=  2 t_0 a \sum_{q,\mu}  \sin((q + A(t))a)  n_{q,\mu}, \label{eq:current_expecation_value_app_1}
\end{align}
where $n_{q, \mu} = \bra{\Psi(t)} \hat{c}_{q,\mu}^\dagger \hat{c}_{q,\mu} \ket{\Psi(t)}$ is the number of electrons with crystal momentum $q$ and spin $\mu$ which remains constant throughout the dynamics. We now expand the sinusoidal-function in Eq. (\ref{eq:current_expecation_value_app_1})
\begin{align}
	j(t) =  2 t_0 a \sum_{q,\mu} n_{q,\mu} &[\sin(qa)\cos(A(t)a) \nonumber \\
	&  + \cos(qa)\sin(A(t)a)]. \label{eq:current_expecation_value_app_2}
\end{align}
We see that the first term in Eq. (\ref{eq:current_expecation_value_app_2}) vanishes as $\Sigma_{q, \mu} \sin(qa) n_{q,\mu} = 0$ for all distributions symmetric around $q=0$ which includes the ground state for $L=10$. Taking a field of constant amplitude, $A(t)=A_0 \sin(\omega_L t + \phi)$, we then expand the sine-function in Eq. (\ref{eq:current_expecation_value_app_2}) using the Jacobi-Anger expansion and obtain
\begin{align}
	j(t) =& 4 t_0 a \sum_{q, \mu}  n_{q,\mu}  \cos(qa) \nonumber \\
	&\times \left( \sum_{n=1}^\infty J_{2n-1}(A_0 a) \sin[(2n-1)(\omega_L t + \phi)] \right), \label{eq:current_expecation_value_app_3}
\end{align}
where $J_n$ is the $n$'th Bessel function of the first kind. The spectrum is finally obtained from the norm square of the Fourier-transform of Eq. (\ref{eq:current_expecation_value_app_3})
\begin{widetext}
\begin{align}
	\tilde{j}(\omega) &= 4 t_0 a \sum_{q, \mu}  n_{q,\mu}  \cos(qa)  \sum_{n=1}^\infty  J_{2n-1}(A_0 a ) \int_{-\infty}^\infty \e^{-i \omega t} \sin[(2n-1)(\omega_L t + \phi)] dt \nonumber \\
	&= -4i \pi t_0 a  \sum_{q, \mu}  n_{q,\mu}  \cos(qa)  \sum_{n=1}^\infty  J_{2n-1}(A_0 a )  \big \{ \e^{i(2n-1)\phi} \delta[\omega - (2n-1)\omega_L] - \e^{-i(2n-1)\phi} \delta[\omega + (2n-1)\omega_L] \big\}, \label{eq:current_fourier_transform_explicit_appendix}
\end{align}
\end{widetext}
where we have assumed an infinitely long pulse for simplicity. We see that Eq. (\ref{eq:current_fourier_transform_explicit_appendix}) is only nonvanishing for odd integer values of $\omega$ showing that only odd harmonics are found in the spectrum in the uncorrelated phase. For a finite pulse (also with an envelope), still only odd harmonics will be found in the spectrum, though less sharply peaked. In that case, the equations become more involved and are not included here.

\bibliography{this_bib_file}

\end{document}